# A SURVEY OF MULTIMEDIA STREAMING IN WIRELESS SENSOR NETWORKS: PROGRESS, ISSUES AND DESIGN CHALLENGES

[1]Taner Cevik, [2]Alex Gunagwera, [3]Nazife Cevik

[1,2]Department of Computer Engineering, Fatih University
Istanbul, Turkey

[3]Department of Computer Engineering, Arel University
Istanbul, Turkey

*ABSTRACT*

*Advancements in Complementary Metal Oxide Semiconductor (CMOS) technology have enabled Wireless Sensor Networks (WSN) to gather, process and transport multimedia (MM) data as well and not just limited to handling ordinary scalar data anymore. This new generation of WSN type is called Wireless Multimedia Sensor Networks (WMSNs). Better and yet relatively cheaper sensors – sensors that are able to sense both scalar data and multimedia data with more advanced functionalities such as being able to handle rather intense computations easily - have sprung up. In this paper, the applications, architectures, challenges and issues faced in the design of WMSNs are explored. Security and privacy issues, over all requirements, proposed and implemented solutions so far, some of the successful achievements and other related works in the field are also highlighted. Open research areas are pointed out and a few solution suggestions to the still persistent problems are made, which, to the best of my knowledge, so far haven't been explored yet.*

## 1.INTRODUCTION

Multimedia (MM), generally, is content that uses a combination of different forms of data such as text, audio, still images, video and/or animations. Enabling WSNs [1] to support MM data has recently become an active focus area for researchers all over the world. This is because of the availability of cheaper CMOS cameras and microphones: hardware which is rather sufficient to make this possible. Successfully achieving this goal, however, is no easy task: with it comes challenges and tough decisions to be made since there are a lot of trade-offs to consider. Like WSNs, WMSNs have a great deal of requirements – some similar to those in WSNs others more complex. In the long run, ideal WMSNs are supposed to be able to sense, retrieve, store, process, transmit and communicate, if need be, scalar data (ie. temperature, humidity, etc.) as normal WSNs [2]-[3] plus still images, audio and video data (MM data). This new sparking opportunity has also posed new challenges to be strived for (i.e. to meet Quality of Service (QoS), bandwidth, time restrictions among other demands required for MM data).

Owing to remarkable advancement in other related research fields such as embedded systems, computer networks, to mention but a few, advanced hardware ranging from cameras, sensor nodes, boards and the like have been manufactured. These have enabled the tackling of even more complicated problems and projects today. These problems and projects come from all sorts of application areas such as the military, health institutions, universities and other academic facilities etc.

In this paper, the prominent applications, general architecture, challenges and issues faced in the design of WMSNs are explored. The rest of this literature is organized as follows. Section II presents an overview of the original communication (protocol) stack – design goals back in the day, brief issues and challenges faced due to requirements for WMSNs today. Section III covers the challenges and issues associated with WMSNs generally. Section IV provides suggested solutions to some of the major issues ranging from security and privacy, routing to QoS problems and issues. Section V presents the available commercial and experimental MM sensors developed. Section VI covers recent, general efficiency and optimization studies. Following, section VII gives classic and new application areas of WMSNs that resulted from technological advances and progress. Lastly, section VIII concludes the paper.

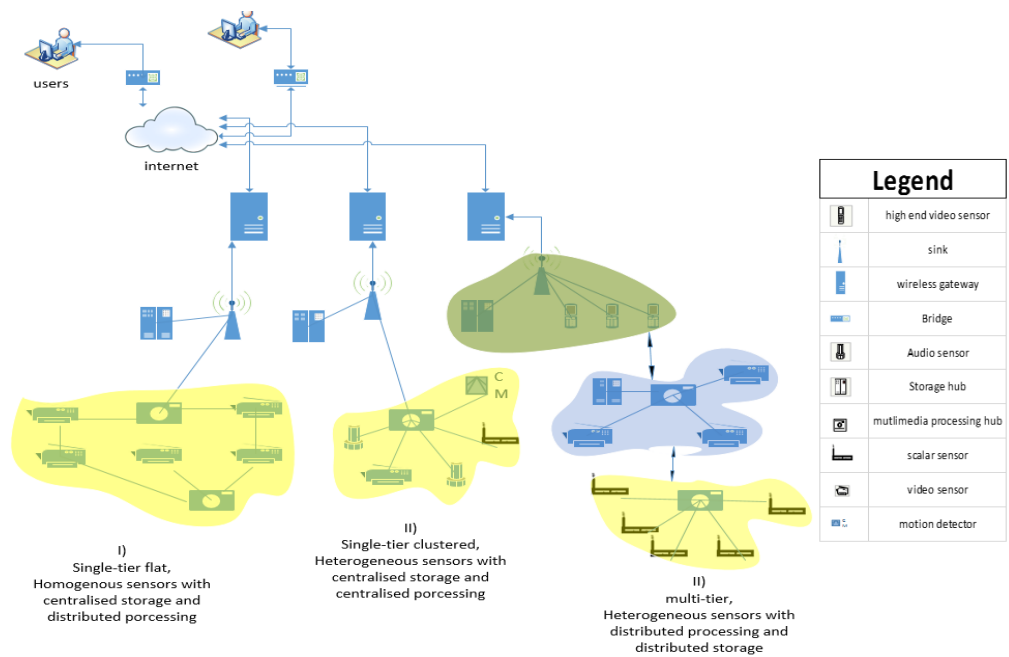

Figure 1. Overall reference architecture of a typical WMSN

## 2. OVERVIEW OF THE PROTOCOL STACK (TRADITIONAL DESIGN GOALS AND EXPECTATIONS)

A brief overview of the requirements and problems faced or likely to be faced at each layer of the stack is clarified in this section. For explanatory purposes, video data will be considered in the subsequent sections till further notice since audio, text and other scalar data are less demanding than video. Thus, if a solution works for videos, then it is most likely to work for audio and other scalar data as well.

### 2.1. MEDIA ACCESS CONTROL (MAC) LAYER REQUIREMENTS

Ordinary MAC Layer protocols designed for conventional wireless networks mainly serve the purposes such as bandwidth allocation, power optimization and awareness, collision prevention, and interference minimization [4]. However, MM streaming requires further struggle and care in order to handle the circumstances specific to MM data transmission, such as optimizing packet latency to meet the end to end delay restrictions and also to provide priority to packets with varying service requirements whilst conserving redundant energy consumption which is the major concern cautiously cared about by the researchers during the design stage of a method or protocol for WSNs.

Suggested MAC schemes specific for WMSNs can be categorized into the following:

- Contention-free schemes: These generally are based on Time Division Multiple Access (TDMA) [5], Code Division Multiple Access (CDMA) [6] and Frequency Division Multiple Access (FDMA) [7].
- Contention-based schemes: Many contention based algorithms for MAC have been suggested, especially for ad hoc networks [8], [9], [13], among others. Since wireless medium is very similar, many of these can be used in WSNs as well. However, IEEE 802.11 is the most widely used since it even has extensions for service differentiation at the MAC level [10].
- Hybrid Schemes: These combine the above two approaches. During the reservation period the sensor nodes in a neighborhood contend for transmission rights and transmission time in the transmission period, based on the amount of data they have to transmit. Once given access to some slots in the transmission period, the sender and the receiver can start to communicate [11-12].

## 2.2. THE NETWORK LAYER

The protocols or methods designed for the network layer is very crucial especially for the provision of QoS in WMSNs because of the reasons as stated below:

- Responsibility for ensuring reliable and energy efficient end to end routes that are used to transmit data.
- Handling the exchange of information between the application layer and MAC layer that affects the communication performance seriously.
- Approaches to improve QoS for MM at the network layer can mainly be grouped into two: timeliness assurance and reliability concerning methods.

**1)Timeliness:** Timeliness concept is concerned for live streaming and applications such as video conferencing, which tend to tolerate a few errors and losses. These time considering methods are broadly grouped into three categories:

- **No priority:** Here the real time packets have got greater priority than the best-effort traffic packets but all real time packets have got the same priority. In [14], an energy aware QoS protocol was proposed. In this method, different packets may have different priorities and the algorithm for calculating multiple paths requires full knowledge of the network at each and every node. This is not only difficult to maintain but also becomes very difficult to scale as the network grows – which it does.

- **Static priority:** All packets have got the same priority for a definite time interval. A static priority routing protocol named SPEED was proposed in [15] to provide real time guarantees for communication in a sensor network. The main problem with this method is that since the priority of packets cannot change, it is difficult to provide the guarantee when miscellaneous changes occur in a network.

- **Differentiated priority:** With differentiated priority mechanism, different priority levels can be assigned to different packet types depending on their inherent characteristics. Felemban et al. [16] presented a differentiated priority packet delivery mechanism called Multi-path Multi SPEED routing protocol (MMSPEED) which is an extension and improvement of the original SPEED [15] architecture.

**2)Reliability:** Applications in this category - such as Video on Demand - are less error tolerant. The commonest method for achieving reliability is the multi-path routing of which the example methods presented in [17-18]. More detailed studies in the area will be provided later in this text. The major drawbacks of ReInForm [17] and MMSPEED [18] respectively are; [17] requires substantial state information to be stored at intermediate sensor nodes and [18] does not consider delay deadlines of the packets while choosing the multiple paths.

## 2.3. TRANSPORT LAYER

Transport layer methods developed for traditional wired and wireless networks are mainly intended for ensuring reliable end to end data (packet) transfer, as well as, to a certain degree, congestion control. However, these issues are not vital for WSNs [19]. This is because sensor nodes are densely deployed, so sensors might be idle with no data transfer most of the time and only become active when an event occurs. Thus, ensuring end-to-end event transfer in most WMSNs makes much more sense and would be more practically efficient than end-to-end packet transfer.

The transport protocols commonly utilized in today's internet (such as TCP, STCP [20] and RTP/RTCP [21]) are not conceivable for WSNs regarding to the following reasons:

- They are unnecessarily complex.

- It is assumed that the main cause of losses is congestion, not noise. However, this does not the point for WMSNs as the transmission medium (air) is mainly noisy. Thus, revisions and improvements are inevitable.

Notable approaches and suggestions for the transport layer can be grouped into 3:

- Congestion control mechanisms: These are combinations of hop-by-hop, flow control, rate-limited traffic and prioritized MAC, the so-called FUSION [22]. The main disadvantage is that it does not support multiple packet priorities, implicit notification and is slow. It also assumes that all sensor nodes have some amount of data – which definitely is not always the case. Others include CODA [23] and Siphon [24] whose major drawbacks are; provides low congestion control and no multi packet priorities and dependent on an effective congestion control algorithm respectively.
- Reliable transport mechanisms: Authors employ a window-less block ACK scheme to handle moderately busty traffic in RBC [25]. Another method, RMST [26] provides correct segmentation and guarantees delivery for all packets. Their major disadvantages, however, include the fact that RBC does not study support for bustier and jitter prone MM traffic and RMST provides reliability for all types of packets, which is wasteful.
- Both congestion control and reliable mechanisms: Methods that belong to this category are concerned with the reliable transmission without congestion. An example study that can be categorized under this title is STCP [27] which is basically based on sessions and provides congestion control as well. Another method, ESRT [28] provides event to sink reliability and congestion control through the use of reporting the rate. The major drawbacks of these methods (STCP and ESRT) include not being scalable and not considering multiple priorities respectively.

## 2.4. CROSS LAYER OPTIMIZATION

Cross layer optimization covers the idea of interaction between the layers of the protocol stack in order to provide outstanding overall performance. The idea was first suggested by Schaar and Shankar in [29]. Since providing QoS for all data sensed including MM is a challenging task, it is crucial that all layers act together. The architecture they suggest takes multimedia data (content with its features, the desired quality of services, etc.), various parameters from the other layers, the degree of adaptability, system constraints and node constraints (such as power, supported bandwidth, delay) as inputs into a system that optimizes the utility given all those constraints and gives the cross layer adaptation strategy and utility (ie. video quality, power, system-wide network utilization) as the output.

The overall approaches at this layer include:

- Methods aimed at minimizing power consumption as presented in [30-31].
- Other approaches are those aimed at maximizing the total throughput. Cui et al propose a cross layer optimized architecture [32] which is concerned with the throughput maximization. Another study of Cui et al [33] which is an extended and further improvement of their previous work, presents that the energy efficiency should be supported across all layers of the protocol stack through a cross-layer design rather than handling independently at each layer.

These approaches' major limitations are that they are either centralized or partially distributed. What is really needed are scalable distributed schemes that require less energy and message exchange. Obviously, an ideal cross layer optimization scheme should be energy aware and have a major purpose of improving QoS.

## 2.5. APPLICATION LAYER

One significant factor that caused the application layer to be one of the most demanding sections of the protocol stack is the process of encoding video, image and/or audio data. This is sometimes computation-intensive depending on the encoding and compression method employed.

Below are some of the main requirements that should be met at the application layer.

- Video, sometimes audio and image coding-compression should be as less complex as possible.
- They should yield low output bandwidth.
- For real time MM streaming, the layer should be designed as loss tolerant.

Recent off the shelf sensor devices are equipped with more energy supplies with embedded microprocessors, enhanced RAM and memory (generally flash), more powerful CPUs and (some very high definition) video capturing cameras as will be discussed in details subsequently.

With all these processing capabilities, video sensors still find it quite challenging to implement motion estimation and compensation techniques based on predictive coding techniques such as those used in the (Moving Picture Experts Group) MPEGx or H.26x series [34]. This necessitates video sensors to employ compression techniques that are founded on coding mechanisms for individual still images, say, the Joint Picture Experts Group (JPEG) or JPEG 2000[35]. Some techniques, such as those in [34] and [36], just avoid motion compensation and/or estimation altogether and instead, encode videos as sequences of images. These are the so called single layer techniques.

Efforts have been made to reduce energy consumption and simplify calculations of the compression techniques of H.26x and MPEGx based on compensation and motion estimation to make them more suitable for WSNs [37-38].

In principle, any combination of two coding paradigms and three compression techniques is possible. However, recently, distributed source coding was investigated in the context of single-layer coding and individual source coding has been explored in the context of three compression techniques used as encoding/compression schemes at the application layer.

To ensure that source coding supports transmission with minimum channel error, joint energy optimization of source code and channel coding research has been carried out as well.

**1) Coding Patterns**

Mainly two coding patterns have been investigated so far for WMSNs. These are the individual source coding and distributed source coding.

- Distributed Source Coding (DSC): DSC refers to the compression of multiple sensor outputs from sensors with limited cooperation and joint decoding at the base station. A more generalized setting for this problem is the multi-terminal source coding problem, also known as the Central Estimating Officer problem [39]. It applies a many-to-one coding paradigm, with the general idea being that the sensors only require less resources to send simply encoded data whereas the decoders are rather complex. Sometimes, in WMSNs, it inevitably happens to be the case that images and/or image frames are captured by MM sensors with overlapping fields of views, a DSC scheme to handle such cases was proposed in [40], in [41], the so-called PRISM (Power efficient Robust hIgh compression Syndrome-based MM coding) was proposed. The authors claim that it is an error resilient DSC scheme. Its greatest advantage is that it reduces amount of data sent by sending only common overlapping image only once using one sensor and other sensors send what is called the "coded syndrome" thereby saving energy. A major drawback of this study is that encoders (senders) should be synchronized since the decoder uses time correlated information to decode the received packets. This requires knowledge of the correlation model of the data from the many senders and this knowledge is hard to obtain.

- Individual Source Coding (ISC): With ISC, each node handles its own encoding/compressing and transmission. It is more popular than DSC since it is much simpler to implement and communication between sensors is not necessary. A major drawback due to this scheme in MWSNs is that it leads to a lot of redundancies since every node must ensure that high quality packets are delivered which causes redundant boost in the number of packets with the lack of information quality.

**2) Layered Coding**

Another way of categorizing the coding techniques is the layered coding as follows: Single Layered Coding, Multi-layered coding and Multiple Description Coding (MDC) that will be briefly examined in this section. Especially a brief overview of some of the most common techniques will be identified later.

- Single Layered Coding (SLC): The most widely used SLC scheme representative is the JPEG with change in which a reference frame is generated and transmitted to the base station periodically. In the time interval period between any two reference frames, sensors only send the differences between the frame and the captured event. However, utilizing fixed point arithmetic rather than floating point in WSNs because of the energy scarcity is a much more promising solution. An example of this study [34] is proposed by Chiasserini et al. in which the JPEG with fixed point Discrete Cosine Transform (DCT) is employed instead of the traditional float point DCT to reduce complexity. An experimental analysis work is performed by Pekhteryev [42] using JPEG 2000 and JPEG on a ZigBee network. Advantages of the scheme include; low processing is required, less total transmission whereas disadvantages include; no resilience or error control nor joint source channel coding, no aggregation either.

Besides reducing the amount of data, another important point is the error concealment during data transmission that is the concept of well resilience source coding. Error concealment can be solved by using Forward Error Correction (FEC) mechanisms or Erasure Correction (EC) codes [43]. However, it is important to note that MM is too large so JPEG may result in faster energy depletion as compared to mechanisms that support aggregation. Also it should be noted that the reference frame is very important because an error in the reference frame will definitely lead to error accumulation throughout all the frames that are encoded with reference to it.

- Multi Layered Coding (MLC): A couple of mechanisms have been proposed but a typical representative schemes proposed are based on the JPEG 2000 [35] and accordingly employ the wavelet transform. One such mechanism [44] clearly shows the trade-off between energy consumption for compression and transmission. It should be noted that there exist general multi-layer compression schemes with motion compensation. However, the investigation of energy efficient motion compensation to exploit temporal redundancy between successive images is still open for research in WMSNs.

Moreover, with the existing multilayer approaches it is difficult to aggregate video data in the vicinity of the event and along the path to the base station. This is because the variable coding and redundancy levels necessitate an adaptive aggregation mechanism, thus requiring both more memory and computing power.

The above two mechanisms have got a serious demerit, it is that if the lower layer is lost at any point, either while on the path due to interference or congestion or noise, the higher layers are rendered completely useless which creates necessity for priorities among layers.

- Multiple Description Coding (MDC): MDC [45] solves the problem as previously mentioned in MLC. Generally, a multiple description coder creates two bit streams of equal importance and the two streams are sent along two separate channels. These two streams all have the same priority. For easy demonstration, let the bit rates of the two channels be R1 and R2 respectively. Thus the total bit rate of the MDC coder would be R1 + R2. The receiving decoder is expected to handle the 2 possible results:

    1. Both R1+R2 arrive: This is handled by a decoder – normally called the central decoder. It gives the highest image/video quality.
    2. If only R1 or R1 arrive, each is received by a different decoder. Both decoders are referred to as side decoders. The both give acceptable quality but, of course, not as efficient as that given by the central decoder.

According to [45] and [46], path diversity along with MDC increases robustness in end to end communication and ensures high bandwidth in radio networks.

Apart from the above mentioned methods, many algorithms specifically for image compression have also been both suggested and developed. Most popular ones will generally be categorized and briefly mentioned in the following. According to [47], they can be generally categorized into two; the *lossy techniques* and the *lossless techniques*.

The lossless image techniques largely depend on two procedures:

1. **Decorrelation:** The stage that removes spatial redundancy between pixels. Then the image compression techniques are applied. The techniques applied after this stage fall under three categories:
    - . Prediction based techniques

- . Transform based techniques
- . Multi resolution based techniques
2. **Entropy encoding:** This second stage is based on the Statistical and Run Length Coding (RLC). The covered lossy JPEG technique for images falls under this stage.

The lossy techniques result in an approximation of the original image. However, these techniques provide a higher compression ratio when compared to the lossless techniques.

## 3. DESIGN CHALLENGES AND PROBLEMS ASSOCIATED WITH WMSNS

Just like WSNs, WMSNs have got requirements that must be met to provide acceptable quality and fulfill inevitable constraints – especially time constraints as will be discussed later. Furthermore, constructing an efficient WMSN requires additional research support from other fields such as signal processing (especially digital signal processing), embedded systems, to mention but a few. This is because MM handling demands much more than what traditional WSNs usually do, thus a more powerful system has to be built. Some of the numerous challenges and issues faced are discussed in this section:

### 3.1. TIME RESTRICTIONS

These actually depend on the type of the application in question, especially important if the application requires live streaming. In this case, end to end delay should be kept as low as possible. Furthermore, in cases where live-streaming is required, more sophisticated algorithms are needed. In applications where live-streaming is not the requirement, sometimes simple methods applying basic techniques such as buffering might suffice, but then again, this might not be feasible for all MM applications as WMSNs have got limited memory with huge amounts of data to process. Thus, buffering might not always work as well.

### 3.2. SECURITY AND PRIVACY

This is a very important and sensitive topic as far as WMSNs are concerned. Some MM applications might demand either security, privacy or even both. Consider a bank-office monitoring application, here things like safe codes should be kept secret, in a parking yards, car number-plates should be kept secret as well. From the security point of view, some ill-intentioned people may just feed useless data to the sensors thus leading to congestion or choking the whole system, which, when worse comes to worst, might even lead to the breakdown of the whole system. Whatever the situation, security, privacy or both might be needed to be ensured – in most cases they are needed.

### 3.3. HUMAN HEALTH AND SAFETY ISSUES

MM sensors should not put humans in peril. This could be directly or indirectly. For example, in industries, radiations from sensors or any incidences of sensor failures and explosions should not ignite gases. Precautions should be made in advance.

Another intriguing issue is whether or not sensors radiations from radio waves actually do cause cancer to humans. The studies in people, especially those who work around radar equipment and those who service communication antennae show no clear increase in cancer risk. Given the exponential increase in number of cell phone usage and users nowadays, studies looking into possible linkages between cell phone usage and cancer were also carried out. Despite one study showing a possible link, most studies did not [48].

## 3.4. QOS REQUIREMENTS

This qualifies to be categorized as one of the most important issues to be dealt with in any WMSN. Vast amounts of research are still being carried out about this topic. Plus, as mentioned before, some applications require real-time streaming/transmission that is live-streaming.

Real time demanding applications can, however, tolerate a few losses. Thus, normally all these requirements are application specific. Hence, a given standard of service must be ensured depending on any given application.

Ways to achieve QoS are mainly through ensuring reliability, timeliness and high quality. Following from the previous section, MMSPEED is very close but it has got its own drawbacks that bar it from being ideal such as the fact that a lot of information needs to be stored in intermediate nodes. Furthermore, it cannot handle network layer aggregation.

## 3.5. LIMITED RESOURCES

As mentioned previously, WSNs are comprised of tiny devices with limited resources. Moreover, MM applications impose additional load on WMSNs in terms of following concepts:

- CPU performance (processing power). This is because MM data is huge in size and requires a lot of operations such as data encoding, decoding thus faster CPUs should be put in place.
- Memory; even though data encoding is done, still the data remains huge and hence, generally, WMSNs end up yearning for much more memory as compared to traditional WSNs.
- Battery power, data rate, among others.

## 3.6. ENERGY CONSUMPTION

For WSNs, energy is a vital resource, which is even of a much more paramount importance for WMSNs if a given standard of QoS is to be ensured. In WMSNs, since very huge computations are carried out, a lot of data processing and transmission of huge data are inevitable. All these operations require energy/power. Unfortunately, one of the most energy consuming operations is radio transmission, which right now is the most common means of transmission in most sensors. Apparently, there is need for energy aware algorithms for WMSNs since the fact that communication functionalities predict power consumption in WSNs does not apply to WMSNs [49].

## 3.7. BANDWIDTH

We have to keep in mind that in wireless networks, bandwidth is not only limited, but also unstable. MM streaming requires very high bandwidth demands to deal with the large amounts of data ranging from scalar to multimedia. High bandwidth with low power spectral density can be provided by Ultra Wide Band (UWB) technology recently.

## 3.8. MM IN-NETWORK PROCESSING

Given the huge size of MM data, schemes such as aggregation would make a very great difference and solve a couple of issues concerned with WMSNs. However, even if aggregation has been successfully applied to scalar data in WSNs, it is very difficult to apply aggregation techniques to WMSNs. This is, thus, one of the open research areas for the future.

## 3.9. CROSS-LAYER OPTIMIZATION

Most of the problems and challenges associated with MM streaming in WSNs differ from stack layer to stack layer. The few tests and even fewer implementations in the field of WMSNs currently are mainly applied at individual stack layers. Difficult as it may be, an efficient – especially energy wise – cross-layer design would be ultimate the best.

## 3.10. RESOURCE ALLOCATION

Given the limited nature of the WMSN resources methods of allocating these resources throughout a network's life time should be put so as to prolong the network's lifespan and to make sure that the network is flexible since networks nowadays actually need to be flexible. Hybrid Automated Repeat Request (HARQ), Schedulers and other functional blocks that operate seamlessly are coupled with Dynamic Resource Allocation (DRA) techniques so as to go about this issue more so for systems that requires flexibility to transmit broad band traffic.

## 3.11. END-TO-END THROUGHPUT

This is responsible for the links in a WMSN – from source to sink regardless of the route used. Currently, there is no protocol specifically designed to serve this purpose yet it is of paramount importance if good performance and QoS is to be achieved. Two protocols were suggested:

- The Radio Link Control protocol (RLP or RLC) [50] - is based on Nacknowledgements (NACK) and does not rely on retransmission timeouts. It is also used in the Universal Mobile Telecommunications Systems (UMTS).
- Adaptive Selective Repeat protocol (ASR) [51] - relies on retransmission timeouts which are configured dynamically.

## 3.12. ROUTING

Routing is another crucial challenging issue to be concerned in the field. Along with it come issues like end to end delay that results from having to transfer huge data and also rather long logged session periods. These need efficient routing algorithms. Routing becomes a complicated issue because;

- Huge amount of data to be concerned with. This in turn creates the need to transport the various packets separately via various paths so as to achieve efficient transmission: the so called *multipath transmission*.
- Since end to end delay is vital in MM applications, especially for real time ones, it is usually preferred to choose the *shortest path* possible as this contributes to the overall QoS and performance of the network. Choosing the shortest path and yet, sometimes, simultaneously considering the multipath factor is not an easy task.
- Sometimes some nodes get overloaded with packets. These nodes need to be successfully bypassed by the subsequent packets. Furthermore, such nodes need to be noticed and bypassed dynamically since routes keep changing from time to time which is challenging to overcome.
- The solutions offered to some of the above problems, discussed in short as follows:
- Use the shortest path possible to reduce the end to end delays.
- Implement less complex algorithms while trying to solve problems regarding routing or when building routing systems.

A lot of research to tackle this topic has been made, but it is still an open area of research. Some of the suggested solutions will be covered later. These include; the adaptive inter spurt approach, letting the source do the transmission path selection among others, etc.

### 3.13. SYNCHRONIZATION

Synchronization can refer to:

- Data/media, i.e. text and audio, video and audio and so on and so forth.
- It could also be network synchronization. Generally one application can employ a vast number of sensors or nodes. These might have varying clocks – which always is the case – due to several reasons ranging from various manufacturers to natural weather phenomena such as temperature. Thus, the need to synchronize these sensors arises as in most cases sensors need to communicate to one another either directly or indirectly or simply share data.

### 3.14. FIELD OF VIEW (FOV)

FOV of a sensor is the angle through which the sensor is sensitive to events. This problem generally arises from the design of the used sensors and in most cases affects the entire WMSN as a whole. Some WMSN applications require that the network to cover a given area completely, say an auto park, a shopping mall, military camp and so forth. However, sometimes it is only required the sensors should be able to focus on a given area when an event occurs. Despite the fact that MM sensors on the market today have got relatively (to classic scalar data sensors) larger FOV and are very sensitive to the direction of data acquisition without even requiring direct LOS between the sensor itself and target object, very few of them are capable of covering the entire area, in fact none. Thus, planning for the entire network and not for individual sensors is much more feasible. Many algorithms have been proposed so far but it is still an open area of research.

### 3.15. COVERAGE

This is concerned with the overall monitoring of a given field. Say an entire battlefield, industrial diagnostics and so on and so forth. Coverage designs of the current data sensors cannot be enough for MM sensors hence there is need for a new model that can support a wider coverage of the MM sensors.

### 3.16. TRANSPORT

Generally, reliable transport is an important issue for WMSNs, however, varies from application to application since some real time streaming applications can tolerate a few loses and some Video on Demand (VoD) applications may not be loss tolerant at all. Currently, there is no specific protocol designed to handle reliable transport issues in WMSNs. SCTP and ESRT mentioned above are quite appealing but they, too, have got their own drawbacks such as not supporting multipath, leading to delay, little or no jitter handling to mention but a few.

## 4. SUGGESTED SOLUTIONS TO THE MAJOR ISSUES

In this section, we will summarize the most promising solutions to the security (for data) challenges, followed by current studies, improvements and solutions to routing and QoS respectively.

The first approach suggested is to permute the position of the data bits. Of course here a known scheme of how to retrieve the data has to be used, [52-53]. However, this method does not guarantee security, yet another approach was suggested which is called *Value transformation* with an idea to transform or reverse the data itself, [54-55]. One that guarantees highest security so far is one that combines the above two suggestions [56].

As mentioned in the previous section, as far as routing is concerned; the shortest path should always be aimed for and used, wherever possible. Also the algorithms used should be as simple as possible. To increase reliability, multi path for MM data should be used.

In order to improve QoS, congestion control is vital. In [57], a study on the various congestion control mechanisms in WSNs, most of which are also applicable to WMSNs as well, is carried out. [58] Proposes a congestion control communication protocol for MM in WSNs. Authors carried out a study on using priorities together with multipath selection for videos in WMSNs [59]. In our previous work [60], we propose a promising multi-channel cross-layer architecture for MM sensor networks with predetermined QoS constraints.

## 5. THE AVAILABLE MM SENSORS, TESTBEDS AND SIMULATORS / EMULATOR    (SOFTWARE AND HARDWARE)

In this section, sample MM sensors, testbeds, simulators and emulators are introduced. Compared to WSNs, MM sensors are still few. However, due to the outstanding progress in technology, many interesting ideas have been brought to life.

Below are some of the available commercial and experimental sensors available today.

### 5.1. STARGATE:

Stargate is manufactured by Crossbow which has an Intel PXA 255 processor that performs at 400Hz. It is also equipped with 32MB of flash memory and 64MB of RAM. It supports computation intensive operations and has got linux OS embedded into it. It also has a 51-pin expansion connector for the MICAz/MICA2 motes and other peripherals such as resolution cameras. It has got a low power consumption. More details can be got from their online datasheet [61].

### 5.2. GARCIA

Even though it is no longer actively maintained or being built, this one was a mobile robot that could be extended with python, C, C++ or even Java. It's equipped with a camera capable of communicating with IEEE 802.4 as well as ZigBee interfaces [62-63].

### 5.3. CYCLOPS

This is a low cost, small size and low power device that supports on-board image processing. Furthermore, it has an image processor and a CPLD as well. It is equipped with an image sensor, micro controller unit and a so called CPLD (complex programmable logic device). According to [64], it can be interfaced with Micaz for communication with other sensor nodes.

### 5.4. WICA

This camera mote has got parallel processor and various camera modules. It is a mote based on Single Instruction Multiple Data (SIMD) and designed for WMSN video analysis processor and a micro-controller (8051 micro controller). WiCa uses the IEEE ZigBee standard for its wireless communication [65].

106

## 5.5. IRISNET

This is a two-tier testbed platform composed of Sensing Agents (SAs) and Organizing Agents (OAs). IrisNet uses aggressive filtering, smart query routing and semantic caching to amazingly reduce bandwidth utilization in the network and also improve query response time.

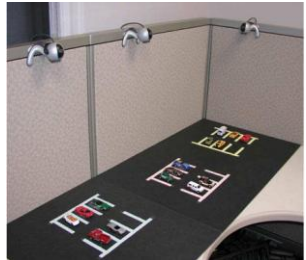

Figure 2. IrisNet platform

Here webcams were used to monitor packing yards for toys. More such demonstrations and details on the platform can be found in [66]. Two main demonstrations about parking space finder and distributed systems monitoring were carried out.

## 5.6. SENSEYE

SensEye is a multi-tier (generally 3-tier) MM sensor network which is also a testbed exceptionally good for surveillance. It is, however, energy intensive and sometimes lacks reliability.

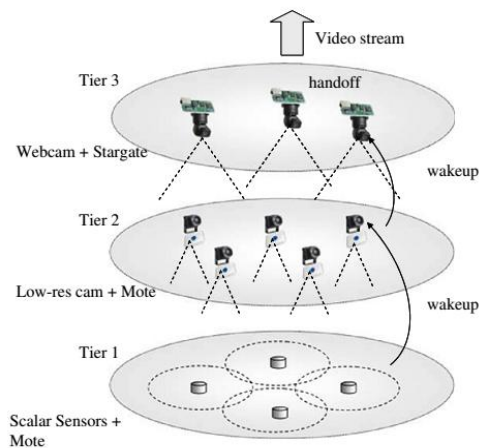

Figure 3. The architecture of a SensEye platform [67]

Figure 3 represents a 3-tier architecture, however 2-tier and 1-tier implementations do exist.

## 5.7. IMOTE2

This sensor has got a low power CPU. It is a low cost, modular sensor built by Intel. It was designed for rather complex advanced applications and supports audio and video-imaging acceleration [68].

## 5.8. MICRELEYE

This is also a low power, low cost sensor. It is equipped with a solar battery and a solar cell. The battery is for emergencies or when the power provided by the cell is not sufficient.

## 5.9. MESHEYE

This was designed for surveillance applications with high energy efficiency as a target. It provides support for in-node processing and multiple resolutions [69-70].

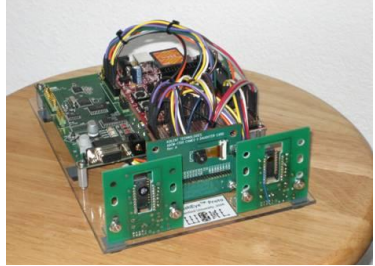

Figure 4. MeshEye mote

## 5.10. CITRIC

It is a camera network system [71] developed that supports in-node processing thereby reducing communication overheads. It has frequency-scalable (up to 624 MHz), 16 MB of flash and 64MB RAM.

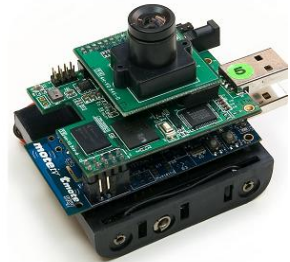

Figure 5. CITRIC

## 5.11. PANOPTES

Panoptes [72] is a low power good quality video sensor platform. It supports video filtering, streaming, compression and capturing.

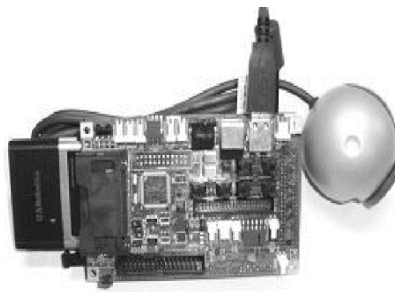

Figure 6. Panoptes

Provides the resolution details, compression, filtering, video capturing performances and many more. However, it consumes about 5watts and has got a video resolution of 320x240. However, the authors claim that the entire device, transmitting over 802.11 consumes about 5.5 watts of power including compression while maintaining the same video resolution and capturing at 18-20 fps.

### 5.12. FOX BOARD BOXED SENSOR

This [73] was developed by acme systems with providing high quality image transmission as the main purpose. Bluetooth is used for transmitting the images captured.

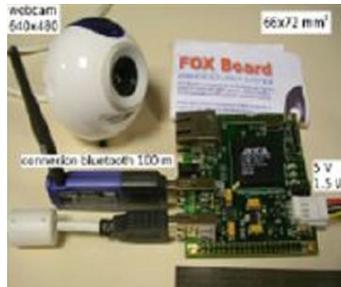

Figure 7. Sensor implemented using the fox board.

## 6. MORE ON EFFICIENCY AND OPTIMIZATION IN WMSNS

A lot of literature has been written on the subject. In this text, we represent but a few of the proposed methods. However, most of the recent proposed methods are along the lines of multi-path streaming of MM and its variations, error concealment/correction. Below is the overview:

In [74], authors suggest error concealment techniques for video transmission over error-prone channels using the new H.26/AV (this is not yet supported by WMSNs –hopefully, sometime later it will) and EC, MDC and Multi-view Video Coding (MVC) techniques.

Using virtual channels leads to a more error-resilient video streaming application [75]. According to this, virtual channels are set up and various packets are sent along those channels. Variations such as assigning priorities to promising channels can also be implemented apparently.

Another interesting idea was to use multi stream coding together with multi path video transporting [76]. This was, however, suggested for ad hoc networks. Whether or not the methods implemented here can be used for WMSNs in general, is still an area open for research.

## 7. APPLICATIONS

WMSNs have not only enhanced applications that use WSNs, but have also led to the realization of some completely new interesting applications resulting from the overtime technological advances. In this section, we cover some of them including some, but not all, applications from [77-78]:

- Gaming: Today WMSNs are applied in games to sense player actions and then stream video/audio. These games can be both online or locally played. Tools such as Wii, Kinect are some of the few practical applications in the field.
- Environment monitoring and control: is very wide area in which WMSNs are used to monitor the environment and gather data. For example, oceanographers use WMSNs to get video feeds on aqua life, wildlife experts and researchers also use HD cameras that either capture moving objects or constantly take pictures of the surrounding and then later review this data for events of interest since it is practically impossible for them to monitor either the parks or the jungles all the time themselves.
- Intelligent homes: This idea is now a reality too. WMSNs have enabled the realization of intelligent homes that enforce both security and enable home customization by using complex

- image and video processing techniques. More specific examples include, identifying cars that belong to the home and then automatically open up the garage doors.
- Health and medicine applications: WMSNs are used not only to monitor patients' bodies or even patients say in mental hospitals but they nowadays also play a role in surgery. Surgeons use gestures to acquire data about the patient's body. This has not only enhanced hygiene but also introduced an efficient way for surgeons to interact with their tools without risking the contamination of the patient or being inconvenienced.

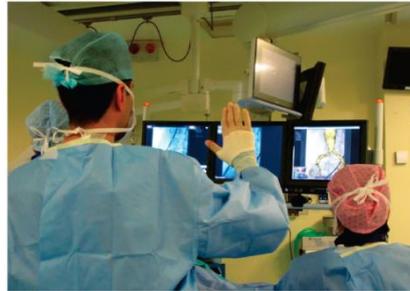

Figure 5. Touchless gesturing in an operation room – an application of WMSNs [79]

- Traffic avoidance, control and enforcement: WMSNs are used to monitor traffic on highways and also avoid traffic congestion by providing routing advice to the users. This has also led to more advanced functionalities such as travelling guidance – it is possible nowadays to travel to and from areas one is completely not familiar with.
- Industrial process control and surveillance: WMSNs are used to monitor machines, check them up for faults and so on. Moreover, they are also used to control temperature and pressure in some processes.
- Law enforcement: This is a general application with examples ranging from shopping malls – to avoid burglars, robbers – to street monitoring cameras to avoid pickpockets and the like. Even if they might not be able to stop the crimes, at the very least they enable the authorities to identify the criminals.

## 8.CONCLUSION

In this study, the general architecture of WMSNs was presented along with challenges and issues associated with achieving efficient, error-resilient and energy aware WMSNs. MM sensors that have come along and those present today with their respective features were also reviewed, previous studies in the field were complemented, open research areas were pointed out, with the present applications of WMSNs already in operation. Prominent solutions to some of the challenges and issues (including the old still persistent one) that have been associated with WMSNs, progress in the field and some of the most promising studies in the area were also presented.

International Journal of Computer Networks & Communications (IJCNC) Vol.7, No.5, September 2015[6] X. Liu, Q. Wang, L. Sha, and W. He, (2003) "Optimal QoS sampling frequency assignment for real-time wireless sensor networks", In Proc. of the 24th IEEE Real-Time Systems Symposium (RTSS), pp308–319.
[7] C. Lu, B. Blum, T. Abdelzaher, J. Stankovic, and H Tian, (2001) "RAP:A real-time communication architecture for large-scale wireless sensor networks", In Proc. of the IEEE Real-time Systems Symposium (RTSS), pp55–66.
[8] Y. Wei, J. Heidemann, and D. Estrin, (2002) "An energy-efficient MAC Protocol for wireless sensor networks", In Proc. of the IEEE INFOCOM, pp1567–1576.
[9] S. Choi, J. Del Prado, S. Nandgopalan, and S. Mangold, (2003) "IEEE 802.11econtention-based channel access (EDCF) performance evaluation", In Proc. IEEE International Conference on Communications (ICC), pp1151–1156, 2003.
[10] M. Adamou, I. Lee, and I. Shin, (2001) "An energy efficient real-time medium access control protocol for wireless ad-hoc networks", 22nd IEEE Real-Time Systems Symposium (RTSS 2001), WIPSession.
[11] V. Rajendran, K. Obraczka, and J. J. Garcia-Luna-Aceves, (2003) "Energy efficient collision-free medium access control for wireless sensor networks", In Proc. of the 1st Int. Conference on Embedded Networked Sensor Systems (SenSys), pp181–192.
[12] X. Yang & N. Vaidya, (2002) "Priority scheduling in wireless ad hoc networks", In Proc. of the 3rd ACM Int. Symposium on Mobile Ad Hoc Networking & Computing (MobiHoc), pp71–79.
[13] K. Akkaya and M. Younis, (2003) "An energy-aware QoS routing protocol for wireless sensor network", In Proc. of the Workshops in the 23rd International Conference on Distributed Computing Systems, pp710–715.
[14] T. He, J. Stankovic, C. Lu, and T. Abdelzaher, (2003) "SPEED: A stateless protocol for real-time communication in sensor networks", In Proc. of the 23rd International Conference on Distributed Computing Systems, pp46–55.
[15] E. Felemban, C-G. Lee, E. Ekici, R. Boder, and S. Vural, (2005) "Probabilistic QoS guarantee in reliability and timeliness domains in wireless sensor networks", In Proc. of the IEEE INFOCOM, pp2646–2657.
[16] E. Felemban, C.-G. Lee, E. Ekici, (2006) "MMSPEED: Multipath multi-SPEED protocol for QoS guarantee of reliability and timeliness in wireless sensor networks", IEEE Transactions on Mobile Computing Vol. 5, No. 6 pp738–754.
[17] B. Deb, S. Bhatnagar, and B. Nath. ReInForM, (2003) "Reliable information forwarding using multiple paths in sensor networks", In Proc. of the 28th Annual IEEE International Conference on Local Computer Networks, pp. 406–415.
[18] C. Wang, M. Daneshmand, B. Li, and K. Sohraby, (2006) "A survey of transport protocols for wireless sensor networks", IEEE Network, Vol. 20, No. 3, pp34–40.
[19] Stream Control Transmission Protocol, http://www.sctp.org.
[20] RTP: A Transport Protocol for Real-Time Applications, http://www.ietf.org/rfc/rfc3350.txt
[21] B. Hull, K. Jamieson, and H. Balakrishnan, (2004) "Mitigating congestion in wireless sensor networks", In Proc. of the ACM SenSys, pp134–147.
[22] C-Y. Wan, S. Eisenman, and A. Campbell, (2003) "CODA: Congestion Detection and avoidance in sensor networks", In Proc. of the ACM SenSys, pp266–279.
[23] C-Y. Wan, S. Eisenman, A. Campbell, and J. Crowcroft, (2005) "Siphon: Overload traffic management using multi-radio virtual sinks in sensor networks", In Proc. of the ACM SenSys, pp. 116–129.
[24] H. Zhang, A. Arora, Y. Choi, and M. Gouda, (2005) "Reliable bursty convergecast in wireless sensor networks", In Proc. of the ACM MobiHoc, pp266–276.
[25] F. Stann and J. Heidemann, (2003) "RMST: Reliable data transport in sensor networks", In Proc. of the 1st International Workshop on Sensor Net Protocols and Applications, pp102–112.
[26] Y. Iyer, S. Gandham, and S. Venkatesan, (2005) "STCP: A generic transport protocol for wireless sensor networks", In Proc. of the IEEE International Conference on Computer Communications and Networks (ICCCN), pp17–19.
[27] Y. Sankarasubramaniam, O. Akan, and I. Akyildiz, (2003) "ESRT: Event-to-sink reliable transport in wireless sensor networks", In Proc. of the 4th ACM Int. Symposium on Mobile Ad Hoc Networking & Computing (MobiHoc), pp177–188.
[28] M. Van der Schaar and S. Shankar, (2005) "Cross-layer wireless multimedia transmission: Challenges, principles, and new paradigms", IEEE Wireless Communications Magazine, Vol. 12, No. 4, pp50–58.
[29] Y. Eisenberg, C. Luna, T. Pappas, R. Berry, and A. Katsaggelos, (2002) "Joint source coding and transmission power management for energy efficient wireless video communications", IEEE Transactions on Circuits and Systems for Video Technology, Vol. 12, No. 6, pp411–424.
[30] M. Gastpar and M. Vetterli, (2005) "Power, spatio-temporal bandwidth, and distortion in large sensor networks", IEEE Journal on Selected Areas in Communications, Vol. 23, No. 4, pp745–754.
111

International Journal of Computer Networks & Communications (IJCNC) Vol.7, No.5, September 2015[55] J.-C. Yen & J.-I. Guo, (1999) "A chaotic neural network for signal encryption/decryption and its VLSI architecture", In Proc. of the 10th VLSI Design/CAD Symposium, pp319–322.

[56] H.-C. Chen, J.-I. Guo, and L.-C. Yen, (2003) "Design and realization of a new signal security system for multimedia data transmission", EURASIP Journal of Applied Signal Processing, Vol. 13, pp1291–1305.

[57] A. Ghaffari, (2015) "Congestion control mechanisms in Wireless Sensor networks: A survey", Journal of Network and Computer Applications, Vol. 52, pp101-115.

[58] S.M. Aghdam et al., (2014) "WCCP: A congestion control protocol for wireless multimedia communication in sensor networks", Ad Hoc Networks, Vol. 13, pp516-534.

[59] L.Zhang, H.Manfred, L.Shu et al, (2008) "Multi¬priority Multi-path Selection for Video Streaming in Wireless Multimedia Sensor Networks", Lecture Notes in Computer Science, Vol. 5061, pp439-452.

[60] T. Cevik & A. H. Zaim, (2013) "A Multichannel Cross-Layer Architecture for Multimedia Sensor Networks", International Journal of Distributed Sensor Networks, Vol. 2013, Article ID 457045.

[61] Willow stargate data, http://www.willow.co.uk/Stargate_Datasheet.pdf.

[62] Acroname GARCIA robot specifications, http://www.acroname.com/garcia/garcia.html.

[63] O. Tekdas, V. Isler, J. H. Lim and A. Terzis, (2009) "Using mobile robots to harvest data from sensor fields", IEEE Wireless Communications, Vol. 16, pp22–28.

[64] N. A. Ali, M. Drieberg and P. Sebastian, (2011) "Deployment of MICAz mote for wireless sensor network applications", In Proc. of the IEEE International Conference on Computer Applications and Industrial Electronics (ICCAIE), pp303–308.

[65] R. Kleihorst, A. Abbo, B. Schueler, A. Danilin, (2007) "Camera mote with a high-performance parallel processor for real-time frame-based video processing", In Proc. of the IEEE Conference on Advanced Video and Signal Based Surveillance (AVSS 07), pp69–74.

[66] S. Nath, Y. Ke, P. B. Gibbons, B. Karp and S. Seshan, (2004) "A distributed filtering architecture for multimedia sensors", 1st Workshop on Broadband Advanced Sensor Networks (BaseNets).

[67] P. Kulkarni, D. Ganesan, P. Shenoy and Q. Lu, (2005) "SensEye: A multi-tier camera sensor network", In Proc. of the 13th annual ACM international conference on Multimedia, pp229–238.

[68] Cross bow Imote2 specifications, http://bullseye.xbow.com:81/Products/Product_pdf_files/Wireless_pdf/Imote2_Datasheet.pdf.

[69] S. Hengstler, D. Prashanth, S. Fong and H. Aghajan, (2007) "MeshEye: A hybrid-resolution smart camera mote for applications in distributed intelligent surveillance", In Proc. of the Information Processing in Sensor Networks, pp360-369.

[70] Stanford MeshEye Mote, http://wsnl.stanford.edu/smartcam.html.

[71] P. Chen, P. Ahammad, C. Boyer, S.I. Huang, L. Lin, E. Lobaton, M. Meingast, S. Oh, S. Wang, P. Yan, A. Yang, C. Yeo, L.C. Chang, J. Tygar, S. Sastry, (2008) "CITRIC: A low-bandwidth wireless camera network platform", In Proc. of the 2nd ACM/IEEE International Conference on Distributed Smart Cameras, pp1–10.

[72] W. Feng, E. Kaiser, W.C. Feng and M.L. Baillif, (2005) "Panoptes: Scalable low-power video sensor networking technologies", ACM Transactions on Multimedia Computing, Communications, and Applications, Vol. 1, No. 2, pp151-167.

[73] Fox boards specifications, http://www.acmesystems.it/hardware_reference.

[74] Z. Cui, Z. Gan, X. Zhan, X. Zhu, (2012) "Error Concealment Techniques for Video Transmission over Error-prone Channels: A survey", Journal of Computational Information Systems, Vol. 8, No. 21, pp8807-8818.

[75] R. Chakravorty, S. Banerjee, S. Ganguly, (2011) "MobiStream: Error-resilient Video Streaming in Wireless WANs using Virtual Channels", In Proc. of the 25th IEEE International Conference on Computer Communications.

[76] S. Mao, S. Lin, S. Panwar, (2003) "Video Transport Over Ad Hoc Networks: Multistream Coding With Multipath Transport", IEEE Journal on Selected Areas in Communications, Vol. 21, No. 10, pp1721-1737.

[77] J.N. Al-Karaki & A. E. Kamal, (2004) "Routing techniques in wireless sensor networks: A survey", IEEE Wireless Communications, Vol. 11, pp6-28.

[78] I.F. Akyildiz, W. Su, Y. Sankarasubramaniam, and E. Cayirci, (2002) "A survey on sensor networks", IEEE Communications Magazine, Vol. 40, pp102-114.

[79] K. O'Hara, G. Gonzalez, G. Penney, A. Sellen, R. Corish, et al, (2014) "Interactional Order and Constructed Ways of Seeing with Touchless Imaging Systems in Surgery", Journal of Computer Supported Cooperative Work, Vol. 23, No. 3, pp299-337.
113

**Authors**

**Taner Cevik** received the B.S., M.S. and Ph.D. degrees in computer engineering from Fatih University in 2001, Fatih University in 2008, and Istanbul University in 2012 respectively. In 2006, he joined the Department of Computer Engineering, Fatih University as a research assistant, and in 2010 became an instructor at the same university. Since 2013, he has served as an assistant professor at Fatih University. 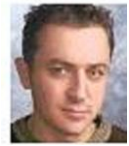

**Alex Gunagwera** received the B.S degree in Computer Engineering and Mathematics from Istanbul Fatih University, in 2014. Followingly, he joined the Department of Computer Engineering, at Fatih University as a research assistant and still proceeds with his M.S. studies. 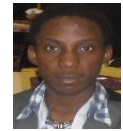

**Nazife Cevik** received the B.S., M.S. and Ph.D. degrees in computer engineering from Fatih University in 2007, 2009, and Istanbul University in 2015 respectively. Since 2015, she has served as an assistant professor at Istanbul Arel University.